\documentclass[10pt,journal]{IEEEtran}

\usepackage{amssymb}
\usepackage{amsmath}
\usepackage{cite}
\usepackage{url}
\usepackage{xcolor}
\usepackage{cite,graphicx,amsmath,amssymb}
\usepackage{fancyhdr}
\usepackage{mdwmath}
\usepackage{mdwtab}
\usepackage{caption}
\usepackage{amsthm}
\usepackage{setspace}
\usepackage{hyperref}
\usepackage{algorithm}
\usepackage{algorithmic}
\usepackage{pdfpages}
\usepackage{mathtools}
\usepackage{subcaption}
\usepackage{balance}
\hypersetup{colorlinks=false}

\graphicspath{{./src/img/}}

\newtheorem{remark}{Remark}
\newtheorem{theorem}{Theorem}

\newtheorem{lemma}{Lemma}

\newtheorem{corollary}{Corollary}

\allowdisplaybreaks
\title{\LARGE Near-Field Velocity Sensing and Predictive Beamforming}

\author{

        Zhaolin~Wang,~\IEEEmembership{Graduate Student Member,~IEEE,}
        Xidong~Mu,~\IEEEmembership{Member,~IEEE,} \\
        and Yuanwei~Liu,~\IEEEmembership{Fellow,~IEEE}
\thanks{Zhaolin Wang is with the School of Electronic Engineering
and Computer Science, Queen Mary University of London, London E1 4NS, U.K. (e-mail: zhaolin.wang@qmul.ac.uk).}
\thanks{Xidong Mu is with the Centre for Wireless Innovation (CWI), Queen's University Belfast, Belfast, BT3 9DT, U.K. (e-mail: x.mu@qub.ac.uk).}
\thanks{Yuanwei Liu is with the Department of Electrical and Electronic Engineering, The University of Hong Kong, Hong Kong (e-mail: yuanwei@hku.hk).}
}

\begin{document}

\maketitle

\begin{abstract}
    The novel concept of near-field velocity sensing is proposed. In contrast to far-field velocity sensing, near-field velocity sensing enables the simultaneous estimation of both radial and transverse velocities of a moving target in the mono-static setup. A maximum-likelihood-based method is proposed for jointly estimating the radial and transverse velocities from the echo signals. Assisted by near-field velocity sensing, a predictive beamforming framework is proposed for a moving communication user, which requires no channel estimation and prior knowledge of user motion model. Finally, numerical examples validate the proposed approaches.
\end{abstract}

\section{Introduction}

In recent years, the exploration of communication and sensing within the near-field region of antenna arrays has garnered significant interest \cite{cui2022near, liu2023near_tutorial, zhang2022beam, friedlander2019localization, d2022cramer, wang2023cram, wang2023near, guerra2021near, 10288339, palmucci2023two, 10380596}, driven by the expanding reach of the near-field region as array apertures and carrier frequencies increase. Unlike the far-field region, where signal propagation is typically planar, the near-field region is characterized by spherical wave propagation. This shift introduces greater complexity in communication and sensing channels due to their reliance on both propagation direction and distance. However, it also opens up opportunities for novel communication and sensing capabilities. In particular, near-field communication can enhance signal multiplexing capabilities through beamfocusing and an increase in the multiple-input multiple-output (MIMO) degrees of freedom \cite{cui2022near, liu2023near_tutorial, zhang2022beam}. Despite these advantages, the dual dependence of near-field channels on direction and distance poses challenges in acquiring accurate channel state information (CSI), potentially leading to prohibitive pilot overhead \cite{cui2022near, liu2023near_tutorial}. From the sensing perspective, spherical-wave propagation makes it possible to infer both direction and distance information from the signal emitted or reflected by the targets. Therefore, most of the existing works focused on joint direction and distance estimation in near-field sensing systems \cite{friedlander2019localization, d2022cramer, wang2023cram, wang2023near}. 

In high-mobility scenarios, such as vehicle-to-everything networks, user/target tracking is an efficient method to guarantee stable communication and sensing performance while meeting the critical latency requirement. In this context, several studies have explored near-field user tracking using communication-only protocols \cite{guerra2021near, 10288339, palmucci2023two}, where the user's state is estimated through Kalman filtering techniques applied to pilot signals sent by the user. However, these approaches may experience a notable decline in effectiveness in real-world applications, primarily due to their failure to account for Doppler frequencies in the signal model and the limited number of pilots typically available in communication-only protocols. To address these limitations, the concept of sensing-assisted predictive beamforming, which leverages integrated sensing and communication (ISAC) techniques, has emerged as a promising solution and has been extensively studied in far-field scenarios \cite{liu2020radar, yuan2020bayesian, mu2021integrated, 10304580, 10284917}. This approach eliminates the need for pilot signals, instead predicting the target's state through the estimated target locations and velocities from echo signals reflected by the target. However, in far-field scenarios, the implementation of sensing-assisted predictive beamforming typically necessitates prior knowledge of the target motion model \cite{liu2020radar, yuan2020bayesian, mu2021integrated, 10304580}. This requirement arises because information about transverse velocity, or angular velocity, is often not discernible from echo signals when utilizing only a single sensing node \cite{niu2022rethinking}, making it difficult to predict the user's direction without the knowledge of the motion model.

Against the above background, we explore the potential of near-field propagation in wireless systems and propose the novel concept of near-field velocity sensing. We reveal that the Doppler frequency in the near-field is influenced by both radial and transverse velocities, which enables \emph{full motion status estimation} of the target. Utilizing this insight, we develop a maximum-likelihood approach for near-field velocity sensing. Based on the proposed near-field sensing, we further propose a sensing-assisted predictive beamforming approach for communication, which eliminates the prior knowledge requirements of the target motion model and facilitates seamless data transmission. Furthermore, unlike traditional far-field predictive beamforming, our approach simplifies algorithmic design due to the accessible information on both radial and transverse velocities. Finally, numerical results are provided to validate the effectiveness of the proposed designs. The code is available at: \url{https://github.com/zhaolin820/near-field-velocity-sensing-and-predictive-beamforming}

\section{Near-Field Velocity Sensing} \label{sec:system_model}

\begin{figure}[t!]
    \centering
    \includegraphics[width=0.4\textwidth]{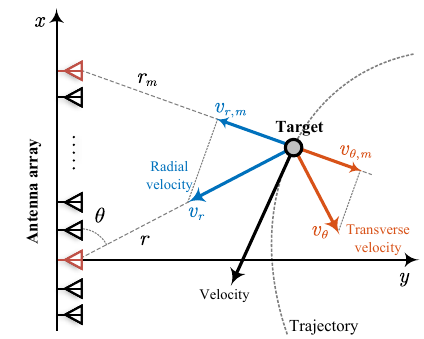}
    \caption{Illustration of the near-field velocity sensing.}
    \label{system_model}
\end{figure}

\subsection{System Model}
As shown in Fig. \ref{system_model}, we consider a narrowband near-field sensing system, which consists of a base station (BS) and a point-like moving target within the near-field region. We assume a shared antenna array for the transmitter and the receiver at the BS, through circulators and full-duplex techniques \cite{6832464}. The bandwidth of this system is denoted by $B$, which corresponds to a symbol duration $T_s = 1/B$.  The BS is equipped with an $M$-antenna uniform linear array (ULA) with spacing $d$. The ULA at the BS is deployed along the $x$-axis and the origin of the coordinate system is put into the center of the ULA. Then, the coordinate of the $m$-th antenna at the BS is given by $(\delta_m d, 0)$, where $\delta_m = m-1-\frac{M-1}{2}$. Let $N$ denote the number of symbol durations for one coherent processing interval (CPI) for near-field sensing, during which the target parameters remain constant. In a specific CPI, let $r$ and $\theta$ denote the distance and angle of the target with respect to the center of the ULA, respectively, and $v_r$ and $v_{\theta}$ denote the \emph{radial velocity} and \emph{transverse velocity} of the target with respect to the center of the ULA, respectively. As such, the coordinate of the target in this CPI is given by $(r \cos \theta, r \sin \theta)$. Let $\mathbf{s}(n) = [s_1(n),\dots,s_M(n)] \in \mathbb{C}^{M \times 1}$ denote the transmit signal of the BS at time index $n$, with $s_m(n)$ representing the transmit signal of the $m$-th antenna. An average power constraint should be satisfied by $\mathbf{s}(n)$, which is given by $\mathbb{E}[ \mathbf{s}^H(n) \mathbf{s}(n) ] = P_t$, with $P_t$ representing the transmit power. The received baseband echo signal at the $m$-th antenna, reflected by the moving target, is given by \cite{tse2005fundamentals, richards2010principles} 
\begin{equation}
    y_{m}(n) = \sum_{i=1}^M \beta_{mi} e^{-j \frac{2\pi}{\lambda} (\bar{r}_m(n T_s) + \bar{r}_i(n T_s) ) } s_i(n) + z_m(n),
\end{equation}
where $\beta_{mi}$ denotes the channel gain, $\bar{r}_m(t)$ denotes the time-variant propagation distance from the $m$-th antenna to the target, $\lambda$ denotes the signal wavelength, and $z_m(n) \sim \mathcal{CN}(0, \sigma^2)$ denotes the complex Gaussian noise.  
In particular, $\bar{r}_m(t)$ can be modeled as $\bar{r}_m(t) = r_m + v_m t$, 
where $r_m$ denotes the distance between the $m$-th antenna at the BS and the target and $v_m$ denotes the velocity component of the target projected along the line connecting the $m$-th antenna at the BS and the target. Based on the geometry relationship illustrated in Fig. \ref{system_model}, $r_m$ and $v_m$ can be expressed as 
\begin{equation}
    r_m = \sqrt{ r^2 + \delta_m^2 d^2 - 2 r \delta_m d \cos \theta  }, \quad v_m = v_{r,m} + v_{\theta,m},
\end{equation}   
where $v_{r,m}$ and $v_{\theta,m}$ are the projections of the radial velocity and the transverse velocity along the line connecting the $m$-th antenna and the target, respectively, and are given by 
\begin{align} 
    \label{velocity_r}
    v_{r,m} = \frac{r - \delta_m d \cos \theta}{r_m} v_r, \quad  v_{\theta,m} = \frac{\delta_m d \sin \theta}{r_m} v_{\theta}.
\end{align} 
Following the radar range equation \cite{richards2010principles}, the channel gain $\beta_{mi}$ can be modeled as $\beta_{mi} = \frac{\beta}{r_m r_i}$. 
Here, $\beta$ is constant within a CPI, satisfying $|\beta|^2 = \frac{G_t G_r \lambda^2 \sigma_{\text{RCS}}}{(4 \pi)^3}$, where $G_t$ and $G_r$ denote the transmit and receive antenna gain, respectively, and $\sigma_{\text{RCS}}$ denotes the radar cross section. The overall received echo signal $\mathbf{y}(n) = [y_1(n),\dots,y_M(n)]^T$ can be expressed as 
\begin{align}
    \mathbf{y}(n) = &\beta \left( \mathbf{A}(r, \theta) \odot \mathbf{D}_n(v_r, v_{\theta}) \right) \mathbf{s}(n) + \mathbf{z}(n),
    \vspace{-0.15cm}
\end{align}      
where $\odot$ is the Hadamard product, $\mathbf{A}(r, \theta) = \mathbf{a}(r, \theta) \mathbf{a}^T(r, \theta)$ with $\mathbf{a}(r, \theta)$ denoting the near-field array response vector, $\mathbf{D}_n(v_r, v_{\theta}) = \mathbf{d}_n(v_r, v_{\theta}) \mathbf{d}_n^T(v_r, v_{\theta})$ with $\mathbf{d}_n(v_r, v_{\theta})$ denoting the Doppler-frequency vector, and $\mathbf{z}(n)$ is the noise vector. More specifically, $\mathbf{a}(r, \theta)$ and $\mathbf{d}_n(v_r, v_{\theta})$ are given by the following equations, respectively:  
\begin{align}
    \mathbf{a}(r, \theta) = &\left[\frac{1}{r_1} e^{-j \frac{2\pi}{\lambda} r_1 },\dots, \frac{1}{r_M}e^{-j \frac{2\pi}{\lambda} r_M } \right]^T, \\
    \mathbf{d}_n(v_r, v_{\theta})  = &\left[ e^{-j \frac{2\pi}{\lambda} v_1 n T_s }, \dots,e^{-j \frac{2\pi}{\lambda} v_M n T_s } \right]^T.
\end{align}

\subsection{Velocity Sensing} \label{velocity_sensing}
Assuming that $\boldsymbol{\eta} = [r, \theta]^T$ have already been estimated\footnote{In Section \ref{sec:predict}, we will show that this assumption is reasonable in near-field velocity sensing systems. }, the remaining unknown parameters are given by $\beta$ and $\boldsymbol{v} = [v_r, v_{\theta} ]^T$. By defining $\mathbf{H}_n(\boldsymbol{\eta}, \boldsymbol{v}) =  \mathbf{A}(r, \theta) \odot \mathbf{D}_n(v_r, v_{\theta})$, the echo signal received over one CPI can be expressed as
\begin{equation}
    \mathbf{Y} = \big[\mathbf{y}(1), \mathbf{y}(2), \dots, \mathbf{y}(N)\big] = \beta \mathbf{X}(\boldsymbol{\eta}, \boldsymbol{v}) + \mathbf{Z},
\end{equation}
where $\mathbf{X}(\boldsymbol{\eta}, \boldsymbol{v}) = [\mathbf{H}_1(\boldsymbol{\eta}, \boldsymbol{v}) \mathbf{s}(1), \dots, \mathbf{H}_N(\boldsymbol{\eta}, \boldsymbol{v}) \mathbf{s}(N)  ]$ and $\mathbf{Z} = [\mathbf{z}(1), \mathbf{z}(2), \dots, \mathbf{z}(N)]$. The unknown parameters $\beta$ and $\boldsymbol{v}$ can be estimated by maximizing the likelihood, which can formulated as the following optimization problem:
\begin{equation} \label{optimization_problem}
    (\hat{\beta}, \hat{\boldsymbol{v}}) = \arg \min_{\beta, \boldsymbol{v} } \left\| \mathbf{Y} - \beta \mathbf{X}(\boldsymbol{\eta}, \boldsymbol{v})  \right\|_F^2,
\end{equation}   
where $\hat{\beta}$ and $\hat{\boldsymbol{v}} = [\hat{v}_r, \hat{v}_{\theta} ]^T$ denote the estimated parameter. For any given $\boldsymbol{v}$, the optimal estimator of $\beta$ is given by
\begin{equation} \label{opt_beta}
    \hat{\beta} = \arg \min_{\beta}  \left\| \mathbf{Y} - \beta \mathbf{X}(\boldsymbol{\eta}, \boldsymbol{v})  \right\|_F^2 = \frac{\mathrm{tr}(\mathbf{Y} \mathbf{X}^H(\boldsymbol{\eta}, \boldsymbol{v}))}{\| \mathbf{X}(\boldsymbol{\eta}, \boldsymbol{v})\|_F^2}.
\end{equation}   
Substituting \eqref{opt_beta} into \eqref{optimization_problem} yields
\begin{equation}
    \left\| \mathbf{Y} - \beta \mathbf{X}(\boldsymbol{\eta}, \boldsymbol{v})  \right\|_F^2 = \|\mathbf{Y}\|_F^2 - \frac{\left|\mathrm{tr}(\mathbf{Y} \mathbf{X}^H(\boldsymbol{\eta}, \boldsymbol{v}))\right|^2}{\| \mathbf{X}(\boldsymbol{\eta}, \boldsymbol{v})\|_F^2}.
\end{equation}
Based on the above result, the optimal estimator of $\boldsymbol{v}$ is
\begin{equation} \label{ML_problem}
    \hat{\boldsymbol{v}} = \arg \max_{\boldsymbol{v}} g(\mathbf{Y}, \boldsymbol{\eta}, \boldsymbol{v}),
\end{equation} 
where 
\begin{equation}
    g(\mathbf{Y}, \boldsymbol{\eta}, \boldsymbol{v}) = \frac{\left|\mathrm{tr}(\mathbf{Y} \mathbf{X}^H(\boldsymbol{\eta}, \boldsymbol{v}))\right|^2}{\| \mathbf{X}(\boldsymbol{\eta}, \boldsymbol{v})\|_F^2}
\end{equation}
is the ML function for estimating $\boldsymbol{v}$. 
Problem \eqref{ML_problem} is an unconstrained optimization problem. Thus, it can be solved by the classical gradient-based methods such as the gradient descent method and the quasi-Newton method, which generally require the gradient of the function $g(\mathbf{Y}, \boldsymbol{\eta}, \boldsymbol{v})$ with respect to the vector $\boldsymbol{v}$. 
According to the chain rule for complex numbers, the gradient $\frac{\partial g(\mathbf{Y}, \boldsymbol{\eta}, \boldsymbol{v})}{\partial v_i}, \forall i \in \{r, \theta\},$ can be expressed as follows \cite{petersen2008matrix}:
\begin{equation} \label{gradient_g}
    \frac{\partial g(\mathbf{Y}, \boldsymbol{\eta}, \boldsymbol{v})}{\partial v_i} = 2\mathrm{Re} \left\{\mathrm{tr}\left( \frac{\partial g(\mathbf{Y}, \boldsymbol{\eta}, \boldsymbol{v})}{\partial \mathbf{X}^T} \frac{\partial \mathbf{X}}{\partial v_i}\right) \right\}.
\end{equation}
Then, we have   
\begin{align}
    &\frac{\partial g(\mathbf{Y}, \boldsymbol{\eta}, \boldsymbol{v})}{\partial \mathbf{X}^T} = \frac{\Theta \mathbf{Y}^H - \Omega \mathbf{X}^H}{\|\mathbf{X}\|_F^4},
\end{align}
where $\Theta = \mathrm{tr}(\mathbf{Y} \mathbf{X}^H)\|\mathbf{X}\|_F^2$ and $\Omega = | \mathrm{tr}(\mathbf{Y} \mathbf{X}^H) |^2$.  
Now, the remaining step is to calculate the partial derivatives $\frac{\partial \mathbf{X}(\boldsymbol{\eta}, \boldsymbol{v})}{\partial v_i} = \left[\frac{\partial \mathbf{H}_1(\boldsymbol{\eta}, \boldsymbol{v})}{\partial v_i} \mathbf{s}(1), \dots, \frac{\partial \mathbf{H}_N(\boldsymbol{\eta}, \boldsymbol{v})}{\partial v_i} \mathbf{s}(N)  \right]$. The expression of $\frac{\partial \mathbf{H}_n(\boldsymbol{\eta}, \boldsymbol{v})}{\partial v_i}$ is derived as follows:
\begin{align}
    \frac{\partial \mathbf{H}_n(\boldsymbol{\eta}, \boldsymbol{v})}{\partial v_i} = &\mathbf{A}(r, \theta) \odot \Bigg( \frac{\partial \mathbf{d}_n(v_r, v_{\theta})}{\partial v_i} \mathbf{d}_n^T(v_r, v_{\theta}) \nonumber \\  & \hspace{1.5cm}+ \mathbf{d}_n(v_r, v_{\theta}) \frac{\partial \mathbf{d}_n^T(v_r, v_{\theta})}{\partial v_i} \Bigg).
\end{align}
The partial derivative $\frac{\partial \mathbf{d}_n(v_r, v_{\theta})}{\partial v_i}$ is given by 
\begin{align}
    &\frac{\partial \mathbf{d}_n}{\partial v_i} = -j \frac{2\pi}{\lambda} nT_s \left[d_{n,1} \frac{\partial v_1}{\partial v_i},\dots, d_{n,M} \frac{\partial v_M}{\partial v_i}   \right]^T,
\end{align} 
where $d_{n,m}$ denotes the $m$-th entry of $\mathbf{d}_n(v_r, v_{\theta})$. According to \eqref{velocity_r}, it can be readily obtained that
\begin{equation} \label{gradient_v}
    \frac{\partial v_m}{\partial v_r} = \frac{r - \delta_m d \cos \theta}{r_m}, \quad \frac{\partial v_m}{\partial v_t} = \frac{\delta_m d \sin \theta}{r_m}.
\end{equation} 
Combining \eqref{gradient_g}-\eqref{gradient_v}, the closed form gradient of function $g(\mathbf{Y}, \boldsymbol{\eta}, \boldsymbol{v})$ with respect to the vector $\boldsymbol{v}$ can be obtained. Then, a stationary point of the ML problem \eqref{ML_problem} can be obtained based on the following gradient ascent process:
\begin{equation}
    \hat{\boldsymbol{v}}^{(t+1)} = \hat{\boldsymbol{v}}^{(t)} + \alpha^{(t)} \left. \frac{\partial g(\mathbf{Y}, \boldsymbol{\eta}, \boldsymbol{v})}{\partial \boldsymbol{v}} \right|_{\boldsymbol{v} = \hat{\boldsymbol{v}}^{(t)}},
\end{equation}
where $\alpha^{(t)} > 0$ denotes the learning rate at the $t$-th iteration. The learning rate at each iteration can be selected using some adaptive method, such as the backtracking line search \cite{armijo1966minimization} or using the existing toolbox, such as the function $\mathtt{fminunc}$ in the optimization toolbox of MATLAB \cite{matlab}. The performance of these methods can be sensitive to the initialization point. Considering the fact that the location and the velocity of the target are not changed significantly between the adjacent CPIs, the initialization point can be selected based on previous CPI estimation to ensure performance, as elaborated in the following section.  

\begin{remark} \label{remark_1}
    \emph{
        \emph{(Benefits of Near-Field Velocity Sensing)} In far-field velocity sensing, since the links between each antenna and the sensing target are approximated to have the same direction, only the radial velocity of the target can be estimated. This fact is further underscored by expressions \eqref{velocity_r}, where $\frac{r - \delta_m d \cos \theta}{r_m} \rightarrow 1$ and $\frac{\delta_m d \sin \theta}{r_m} \rightarrow 0$ as the distance $r$ becomes sufficiently large. However, within the near-field region with relatively small $r$, the simultaneous estimation of both radial and transverse velocities becomes achievable. In this case, the instantaneous motion status of the target can be obtained without the prior knowledge of the target motion model required in far-field regimes \cite{liu2020radar}.
    }
\end{remark}

\section{Predictive Beamforming Through Near-Field Velocity Sensing} \label{sec:predict}

In this section, we explore using near-field velocity sensing in predictive beamforming for near-field communications. Typically, beamforming relies on channel state information influenced by the user's location, necessitating channel estimation before data transmission in each CPI. Despite existing low-complexity methods, the pilot overhead remains high due to near-field channels requiring both distance and angular information between transceivers. To tackle this, we propose leveraging near-field velocity sensing for user tracking, enabling direct retrieval of the user's current location based on prior sensing results. This approach allows for pilot-free and seamless predictive beamforming.

\subsection{Proposed Predictive Beamforming Framework}
The proposed framework consists of the following steps.
\begin{itemize}
    \item \textbf{Initial Access:} 
    This step happens in the first CPI, which does not transmit data and focuses only on sensing. In particular, upon the entry of a mobile user into the coverage area of the BS, the user first informs the BS of its initial location information $\hat{\boldsymbol{\eta}}_0 = [\hat{r}_0, \hat{\theta}_0]^T$ obtained by the classical method, such as beam training or global navigation satellite system, which serves as the initial input of the proposed framework and facilitates estimating $\hat{\boldsymbol{v}}_0 = [\hat{v}_{r,0}, \hat{v}_{\theta,0}]^T$ in the first CPI.
    \item \textbf{Beam Prediction:} Given estimated velocities, the user location in the next CPI can be predicted, which is then used to design the beamforming in the next CPI. 
    \item \textbf{Data Transmission and Velocity Sensing:} Leveraging the predicted user location and the predictive beamformer obtained in the previous CPI, the data can be effectively transmitted to the user and the radial and transverse velocities of the user in the current CPI can be estimated from the echo signals.
\end{itemize}
By repeating the second and third steps, pilot-free and seamless communication between the BS and a mobile user can be achieved in the near-field region. Next, we will elaborate on the communication model and the beam prediction method. 

\subsection{Communication with Predictive Beamforming}
Let $r_l$, $\theta_l$, $v_{r, l}$, and $v_{\theta, l}$ denote the distance, angle, radial velocity, and transverse velocity of the user in the $l$-th CPI, respectively. Then, the user receives the following signal at time $n$ of the $l$-th CPI:
\begin{equation}
    y_{c,l}(n) = \mathbf{h}_l^H(n) \mathbf{s}_l (n) + z_{c,l}(n),
\end{equation}
where $\mathbf{h}_l^H(n) = \beta_c \mathbf{a}^T(r_l, \theta_l) \mathrm{diag}\left( \mathbf{d}_n(v_{r,l}, v_{\theta, l}) \right) \in \mathbb{C}^{1 \times M}$ denotes the communication channel, $\beta_c$ is a constant containing antenna gain and carrier frequency, $\mathbf{s}_l(n) \in \mathbb{C}^{M \times 1}$ denotes the transmit signal satisfying $\mathbb{E}[ \mathbf{s}_l^H(n) \mathbf{s}_l(n) ] = P_t$ and $z_{c,l}(n) \sim \mathcal{CN}(0, \sigma^2)$ denotes the complex Gaussian noise. 
Let $\hat{\boldsymbol{v}}_l = [\hat{v}_{r,l}, \hat{v}_{\theta,l} ]^T$ and $\hat{\boldsymbol{\eta}}_{l+1} = [\hat{r}_{l+1}, \hat{\theta}_{l+1}]^T$ denote the estimated velocity vector and the predicted user location vector in the $l$-th CPI, respectively. In particular, the velocity vector $\hat{\boldsymbol{v}}_l$ can be estimated using maximum likelihood method proposed in Section \ref{velocity_sensing} based on the predicted/initial location $\hat{\boldsymbol{\eta}}_l$ in the $(l-1)$-th CPI, which is given by 
\begin{equation} \label{ML_solution}
    \hat{\boldsymbol{v}}_l = \arg \max_{ \boldsymbol{v}_l } g(\mathbf{Y}_l, \hat{\boldsymbol{\eta}}_l, \boldsymbol{v}_l),
\end{equation}  
where $\mathbf{Y}_l$ denotes the overall echo signal received at the BS in the $l$-th CPI. With $\hat{\boldsymbol{v}}_l$ at hand, the entries of $\hat{\boldsymbol{\eta}}_{l+1}$ can be calculated based on the physical relationship as follows:
\begin{equation} \label{estimated_r_theta}
    \hat{r}_{l+1} = \hat{r}_l + \hat{v}_{r, l} N T_s, \quad
    \hat{\theta}_{l+1} = \hat{\theta}_l + \frac{\hat{v}_{\theta, l} N T_s}{\hat{r}_l},
\end{equation}      
where $NT_s$ denotes the duration of a CPI. After obtaining $\hat{\boldsymbol{v}}_l$ and $\hat{\boldsymbol{\eta}}_{l+1}$, the beamforming in the $(l+1)$-th CPI can be designed to maximize the communication rate. Two key factors should be considered for this objective: \emph{array gain maximization} and \emph{Doppler-frequency compensation}. Specifically, maximizing array gain required the knowledge of user location, which has been predicted through \eqref{estimated_r_theta}. To compensate for the Doppler frequency, the velocity information of the user is needed. Considering that the acceleration of typical mobile users (e.g., passenger vehicles), is generally small, it is safe to use the velocity estimated in the previous CPI to compensate for the Doppler frequency. For example, let us consider a vehicle with an acceleration of $10 \text{ m/s}^2$. The change of its velocity after a CPI of $0.02$ s is only $0.2$ m/s. Therefore, the predictive beamformer at time $n$ for the $(l+1)$-th CPI can be designed as
\begin{equation} \label{predictive_signal}
    \mathbf{w}_{l+1}(n) = \sqrt{\rho_l} \mathrm{diag}( \mathbf{d}_n^*( \hat{v}_{r,l}, \hat{v}_{\theta, l}) )  \mathbf{a}^*(\hat{r}_{l+1}, \hat{\theta}_{l+1}), 
\end{equation}
where $\rho_l$ is the power regularization factor such that $\|\mathbf{w}_{l+1}\|^2 = P_t$. The transmit signal at time $n$ of the $(l+1)$-th CPI is thus given by
\begin{equation}
    \mathbf{s}_{l+1}(n) = \mathbf{w}_{l+1}(n) c_{l+1}(n),  
\end{equation}  
where $c_{l+1}(n) \in \mathbb{C}$ denotes the data symbol that can be modeled as an independent random variable with zero mean and unit power. Then, the average achievable rate over the $l$-th CPI is given by \cite{tse2005fundamentals}  
\begin{equation}
    R_l = \frac{1}{N} \sum_{n=1}^N \log \left( 1 + \frac{1}{\sigma^2}|\mathbf{h}_l^H(n) \mathbf{w}_l(n)|^2 \right).
\end{equation} 
The proposed predictive beamforming framework is summarized in \textbf{Algorithm \ref{alg:predictive}}.

\begin{algorithm}[tb]
    \caption{Proposed Predictive Beamforming Framework.}
    \label{alg:predictive}
    \begin{algorithmic}[1]
        \STATE{\emph{Initial access:} Estimate the velocity $\hat{\boldsymbol{v}}_0$ in the first CPI based on the initial location information $\hat{\boldsymbol{\eta}}_0$ and formulate the beamformer $\mathbf{w}_1(n)$ according to \eqref{estimated_r_theta} and \eqref{predictive_signal}; Set $l = 1$.}
        \REPEAT
        \STATE{Transmit data to the user using beamformer $\mathbf{w}_l(n)$.}
        \STATE{Collect echo signals $\mathbf{Y}_l$. }
        \STATE{Estimate radial and transverse velocities $\hat{\boldsymbol{v}}_l$ by \eqref{ML_solution}.}
        \STATE{Predict the next user location $\hat{\boldsymbol{\eta}}_{l+1}$ by \eqref{estimated_r_theta}.}
        \STATE{Design the next beamformer $\mathbf{w}_{l+1}(n)$ by \eqref{predictive_signal}. }
        \STATE{Set $l = l+1$. }
        \UNTIL{the user moves out of the coverage area of the BS.}
    \end{algorithmic}
\end{algorithm}

\begin{remark}
    \emph{
        \emph{(Simplified Algorithmic Design)} Thanks to the availability of both radial and transverse velocity data, the user's state can be straightforwardly predicted in the simplest way, as shown in \eqref{estimated_r_theta}. Consequently, this eliminates the necessity for complex algorithmic designs for prediction.
    }
\end{remark}

\section{Numerical Examples}
In this section, numerical examples are provided to validate the feasibility of near-field velocity sensing and the effectiveness of the proposed predictive beamforming framework. The BS is assumed to be equipped with $M = 512$ antennas with $\lambda/2$ spacing. The carrier frequency is set to $28$ GHz. The system bandwidth and the length of a CPI are set to $B = 100$ KHz and $2$ ms, respectively. Thus, we have $T_s = 10^{-5}$ s and $N = 200$. Furthermore, we set $G_t = G_r = 1$ and $\sigma_{\text{RCS}} = -23$ dB. The noise density power is set to $-174$ dBm/Hz.

\begin{figure}[t!]
    \centering
    \includegraphics[width=0.4\textwidth]{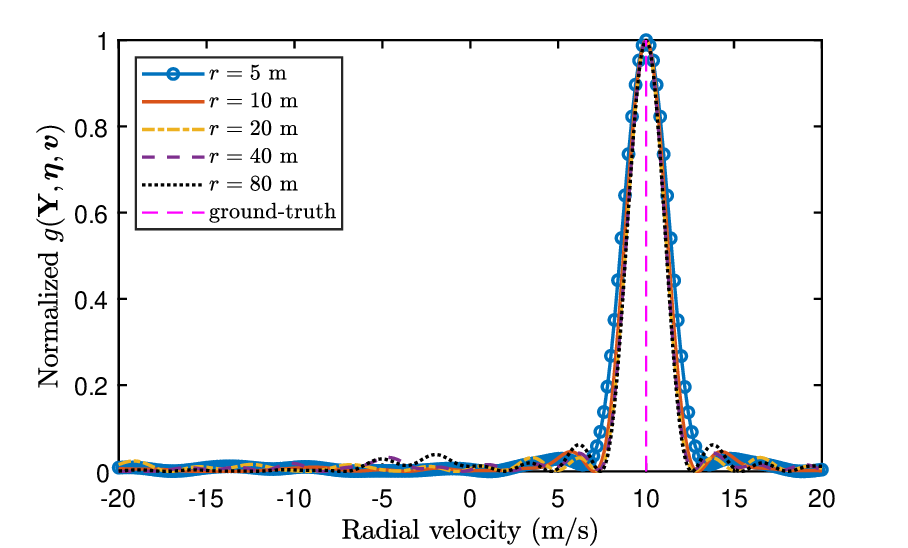}
    \includegraphics[width=0.4\textwidth]{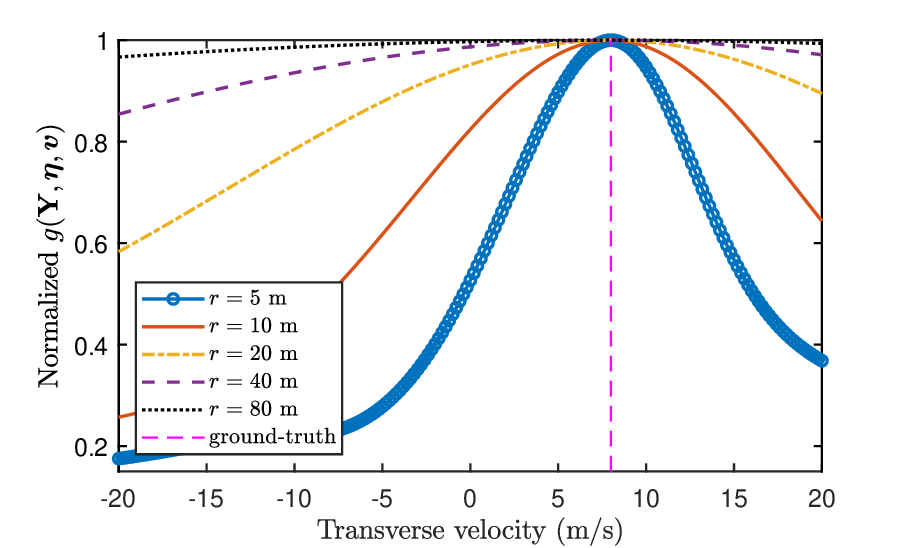}
    \caption{Impact of the distance on the ML function $g(\mathbf{Y}, \boldsymbol{\eta}, \boldsymbol{v})$.}
    \label{fig:sensing_performance}
\end{figure}

\begin{figure}[t!]
    \centering
    \includegraphics[width=0.4\textwidth]{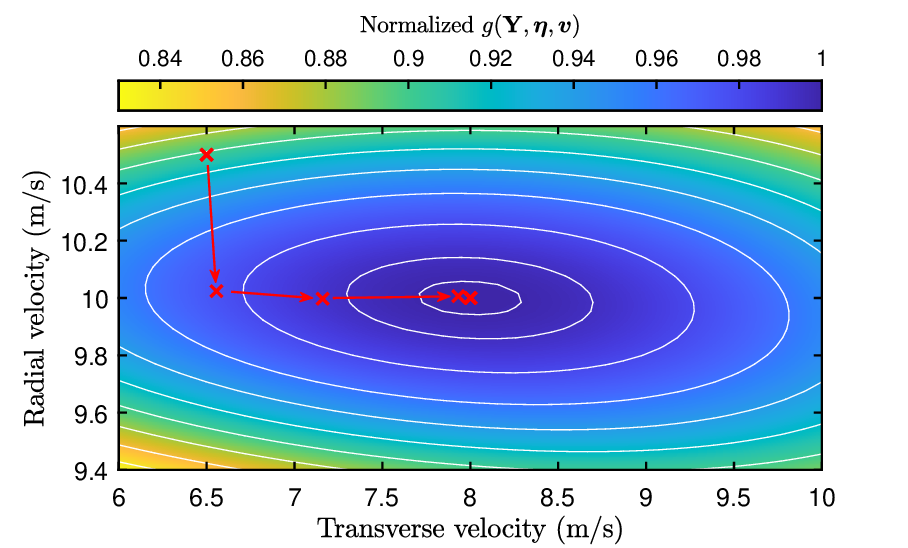}
    \caption{Convergence behavior of the proposed gradient ascent method.}
    \label{fig:convergence}
\end{figure}

In Fig. \ref{fig:sensing_performance}, the impact of the distance parameter on the ML function $g(\mathbf{Y}, \boldsymbol{\eta}, \boldsymbol{v})$ for velocity sensing is illustrated. The radial and transverse velocities of the targets are set to $v_r = 10$ m/s and $v_{\theta} = 8$ m/s, respectively. Other parameters of the target are assumed to be perfectly matched with the ground truth. For a fair comparison, the transmit power is adjusted according to the distance such that the signal-to-noise ratio remains the same. As can be observed from Fig. \ref{fig:sensing_performance}(a), the ML function demonstrates a main lobe in close proximity to the ground truth, regardless of whether the target is near or far. It is also interesting to see that the main lobe becomes slightly narrower, which indicates improved performance in radial velocity sensing, as the target moves toward the far-field region. This phenomenon arises because, with increasing distance, the radial velocity manifested in the echo signal at each antenna gradually becomes the same, which is preferred by the radial velocity sensing. Concerning transverse velocity sensing, as illustrated in Fig. \ref{fig:sensing_performance}(b), it is notable that the ML function also has a main lobe in proximity to the ground truth. However, this main lobe progressively diminishes as the distance $r$ increases. At a distance of $r = 80$ m, the ML function flattens over a substantial interval around the ground truth, rendering transverse velocity sensing impractical. These results are consistent with \textbf{Remark \ref{remark_1}}. The convergence behavior of the proposed gradient ascent algorithm for velocity estimation is depicted in Fig. \ref{fig:convergence}, where the $\mathtt{fminunc}$ function in MATLAB is used to additively select the learning rate. It can be observed that the proposed algorithm can converge to the minimum ground the ground-truth within a few iterations.

\begin{figure}[t!]
    \centering
    \includegraphics[width=0.4\textwidth]{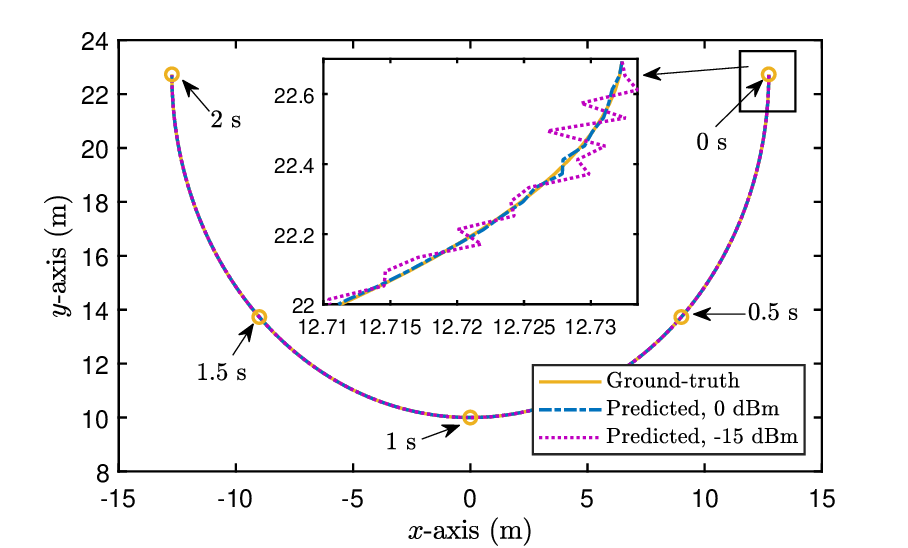}
    \caption{Predicted trajectory of the moving user.}
    \label{fig:trajectory}
\end{figure}

\begin{figure}[t!]
    \centering
    \includegraphics[width=0.4\textwidth]{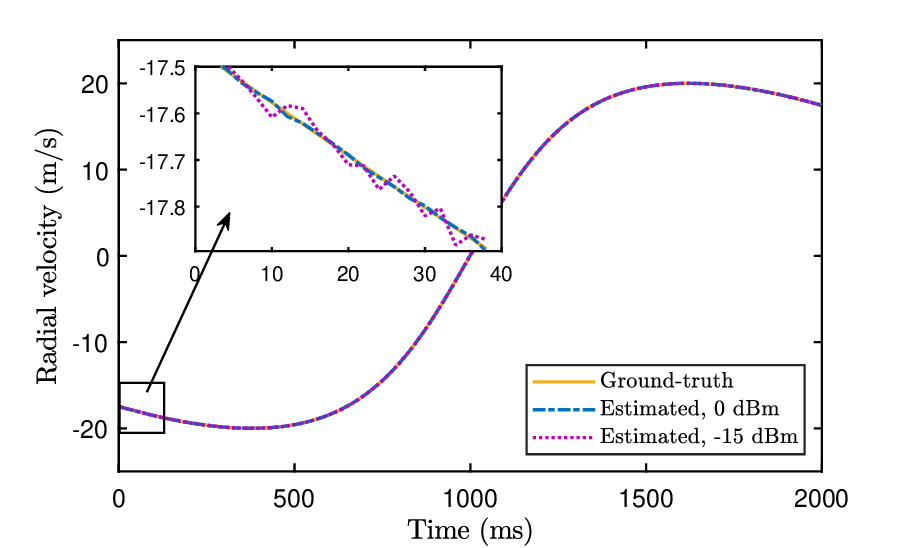}
    \includegraphics[width=0.4\textwidth]{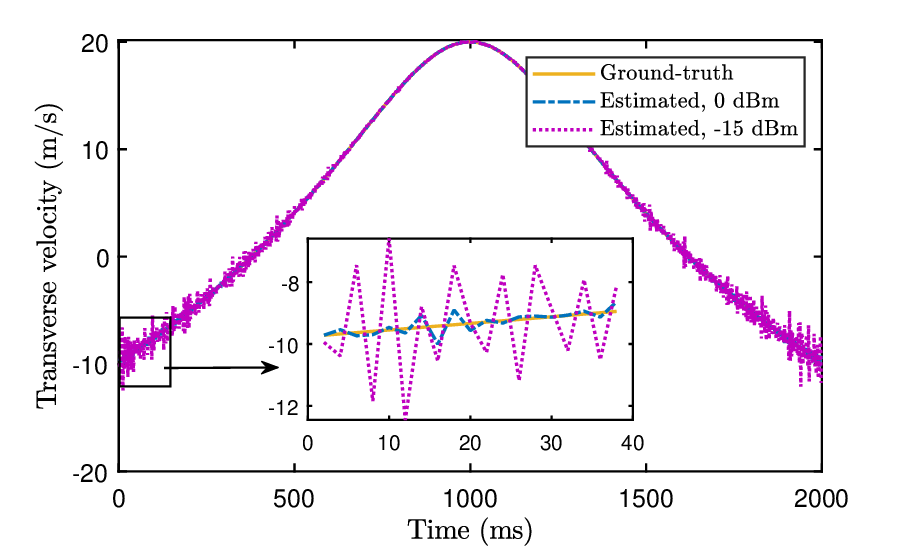}
    \caption{Estimated velocities of the moving user at different times.}
    \label{fig:velocity_sensing}
\end{figure}

Next, we study the performance of the proposed predictive beamforming scheme. Without loss of generality, we consider a moving user following the trajectory depicted in Fig. \ref{fig:trajectory} at a speed of $20$ m/s, with the BS fixed at the origin of the coordinate system. \textcolor{black}{The initial user location and parameter $\beta$ are assumed to be perfectly estimated at the initial access stage}. The predicted trajectories, obtained through the proposed scheme at varying transmit powers, are also visualized in Fig. \ref{fig:trajectory}. Notably, the predicted trajectories closely align with the ground truth owning to the new velocity sensing ability in the near-field region. To gain more insights, Fig. \ref{fig:velocity_sensing} presents the estimated radial and transverse velocities of the moving user at each CPI, both of which closely match the ground-truth values, while transverse velocity sensing is more sensitive to user-BS distance compared to radial sensing. \textcolor{black}{Furthermore, the robust performance of the proposed method is evident as it relies on predicted rather than ground-truth user locations, confirming its resilience against uncertainties.}

\begin{figure}[t!]
    \centering
    \includegraphics[width=0.4\textwidth]{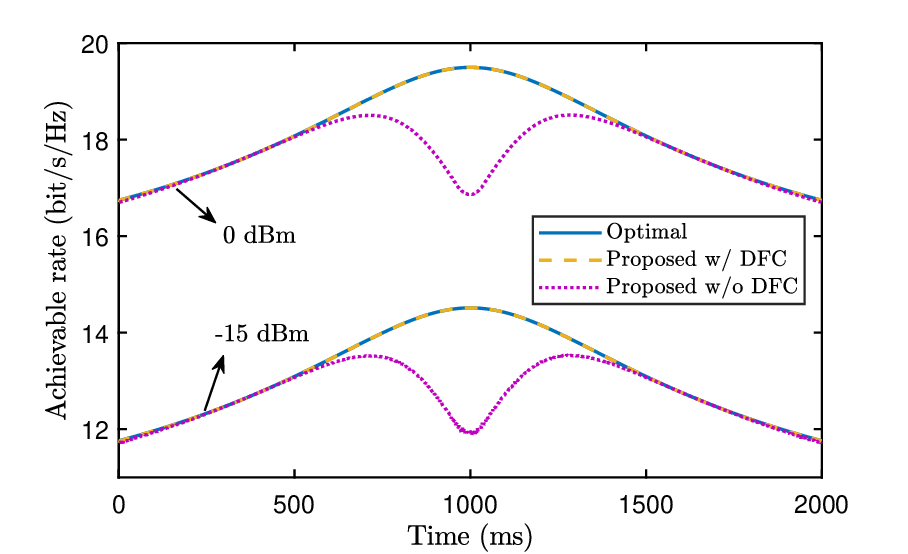}
    \caption{Achievable rate through predictive beamforming.}
    \label{fig:rate}
\end{figure}

Fig. \ref{fig:rate} illustrates the achievable rate $R_l$ attained through the proposed predictive beamforming scheme with and without Doppler-frequency compensation (DFC). For the scheme without DFC, the beamformer in the $(l+1)$-th CPI is designed as $\bar{\mathbf{w}}_{l+1}(n) = \sqrt{\rho_l} \mathbf{a}^*(\hat{r}_{l+1}, \hat{\theta}_{l+1})$. Furthermore, the optimal achievable rate is calculated by utilizing the ground-truth velocities and locations during the beamforming design. The results demonstrate that the proposed predictive beamforming achieves a performance closely approaching the optimal when the Doppler frequency is compensated using the estimated velocities in the previous CPI. However, without the Doppler frequency compensation, there is a significant performance loss when the transverse velocity is dominated, highlighting the necessity of DFC in near-field systems.

\section{Conclusion}
In conclusion, a novel near-field velocity sensing method was introduced, enabling simultaneous estimation of radial and transverse velocities of a moving target. This method facilitated a predictive beamforming approach that maximizes array performance and compensates for Doppler frequency without the need for channel estimation and prior knowledge of the target motion model. Numerical results demonstrated the effectiveness of these approaches, suggesting new possibilities for wireless sensing and communication through near-field propagation. 
Future research directions include extending velocity sensing and predictive beamforming to multi-user or wideband scenarios and developing low-complexity algorithms based on methods such as nonlinear filtering, compressive sensing, and deep learning for near-field velocity sensing.

\balance
\bibliographystyle{IEEEtran}
\bibliography{reference/mybib}

\end{document}